\documentclass{article}
\usepackage{amsmath}
\usepackage[utf8]{inputenc}
\usepackage{amssymb}
\usepackage{graphicx}
\title{A Visualization of Null Geodesics for the Bonnor Massive Dipole}
\author{Oliva-Mercado, G. Andree \& Bonatti-González, Javier  \\ \& Cordero-García, Iván \& Frutos-Alfaro, Francisco. }
\date{2014-2-12}
\begin{document}
\section*{A Visualization of Null Geodesics for the Bonnor Massive Dipole}
\begin{center}
Oliva-Mercado, G. Andree\footnote{Escuela de Física, Universidad de Costa Rica, gandreoliva@gmail.com} \& Bonatti-González, Javier \footnote{Escuela de Física, Universidad de Costa Rica, jbonatti2011@gmail.com} \\ \& Cordero-García, Iván\footnote{Escuela de Física, Universidad de Costa Rica, ivancg@msn.com} \& Frutos-Alfaro, Francisco\footnote{Centro de Investigaciones Espaciales, Universidad de Costa Rica, frutos62@gmail.com}
\end{center}

\setcounter{page}{0}
\hrulefill

\textbf{Abstract}

In this work we simulate null geodesics for the Bonnor massive dipole metric by implementing a symbolic-numerical algorithm in Sage and Python. This program is also capable of visualizing in 3D, in principle, the geodesics for any given metric. Geodesics are launched from a common point, collectively forming a cone of light beams, simulating a solid-angle section of a point source in front of a massive object with a magnetic field. Parallel light beams also were considered, and their bending due to the curvature of the space-time was simulated.

\vspace{1cm}
\hrulefill

\textbf{Resumen}

En este trabajo se simulan geodésicas nulas para la métrica de dipolo masivo de Bonnor, implementando un algoritmo simbólico-numérico en Sage y Python. Este programa es capaz, en principio, de visualizar en 3D las geodésicas de cualquier métrica. Estas geodésicas inician en un punto común, formando colectivamente un cono de rayos de luz, simulando una sección de ángulo sólido de una fuente puntual frente a un objeto masivo con campo magnético. Se consideran también rayos de luz paralelos, y se simuló su cambio de trayectoria debida a la curvatura del espacio-tiempo. 

\vspace{2cm}
\textbf{Keywords:} General Relativity, Geodesics, Differential Geometry, Free Software.

\textbf{Palabras clave:} Relatividad General, Geodésicas, Geometría Diferencial, Software Libre.

\newpage
\setcounter{page}{1}
\maketitle

\section{Introduction}
In General Relativity, Einstein field equations are a set of non-linear partial differential equations that describe the curvature of space-time due to the presence of mass-energy-momentum. One of the exact solutions for the Einstein field equations in the vacuum was proposed in 1965 by W.B. Bonnor\cite{bonnor}; it describes a massive spherical stationary object with a dipole magnetic field. This solution, expressed using the metric line element $ds^2=g_{\mu\nu}dx^\mu dx^\nu$ (\footnote{Einstein summation convention applies.}), in which $g_{\mu\nu}(\mathbf{x})$ is the metric describing the manifold solution to these equations is:

\begin{equation}
 ds^2 = -\frac{Y^2 P^2}{Q^3Z}(dr^2+Z d\theta^2)-Z\left(\frac{Y}{P}\right)^2\sin^2\theta d\phi^2+\left(\frac{P}{Y}\right)^2 dt^2
\end{equation}
where
\begin{subequations}
\begin{eqnarray}
P=r^2-2mr-b^2cos^2\theta \\
Q = (r-m)^2-(b^2+m^2)cos^2\theta\\
Y=r^2-b^2cos^2\theta\\
Z=r^2-2mr-b^2
\end{eqnarray}
\end{subequations}
$t,r,\theta,\phi$ are the standard spherical coordinates. The geometric system of units is used along this paper ($c=G=1$). The parameters $m$ and $b$ are defined, according to Bonnor, as $b$ being such that if $r$ becomes large, the magnetic field is that of a magnetic dipole along $\theta=0$ of moment $m b \pi^{-\frac{1}{2}}$ and the gravitational field becomes that of a mass $2m$ at the origin.\cite{bonnor}

Trajectories on this metric must be found by using the geodesic equations, which can be thought as lines of minimum length (result from $\delta \int ds = 0$):
\begin{equation}\label{eq:geodesicequation}
\frac{d^2 x^\kappa}{d\lambda^2}=-\Gamma^\kappa_{\mu\nu}\frac{dx^\mu}{d\lambda}\frac{dx^\nu}{d\lambda}
\end{equation}
subject to initial conditions that determine the kind of geodesics\footnote{The sign of $ds^2$ that determines kind of geodesic depends on the signature of the metric. In this paper, the signature of the metric is $(1,-1,-1,-1)$}: time-like ($ds^2>0$, trajectories of particles with mass), light-like or null ($ds^2=0$, trajectories of photons), or space-like ($ds^2<0$, trajectories of hypothetical faster-than-light particles). $\lambda$ is an affine parameter, and $\Gamma_{\mu\nu}^\kappa$ are the components of the Christoffel symbol.

In this paper null geodesics for the Bonnor metric are calculated in order to simulate the effect of the curvature of the space-time on a point source of electromagnetic radiation located at a certain distance from the origin, where the massive object is located. To achieve this purpose, a program in Python and Sage was developed to calculate analytically the Christoffel symbols from the metric and use them to build the geodesic equations. Once this is done, the equations are solved numerically using user-specified initial conditions, and these conditions are then varied to simulate a cone of light beams that represents a solid-angle section of the source.

Other efforts have been made in this topic. For example, Müler and Frauendiener (2010, \cite{muller}) developed Geodesic Viewer, which implements a selection of metrics which does not include the one studied here; and unlike this program, a local frame (tetrads) is required to specify the initial direction of the light beam. The justification for not using a local frame is that our scope is to study the \emph{collective} behavior of geodesics, and high precision knowledge of the local directions is not required, as it would be for calculating gravitational lensing for such a point source. The Bonnor geodesics were also previously studied in 2D by Sanabria and Valenzuela \cite{sanabria}, for charged test particles, as well as by Kovář et al. \cite{kovar}, for neutral particles, from the point of view of orbit stability.

\emph{Sage} is an integrated environment consisting of several open-source packages (last count 89) with a common Python-based interface. It is capable of handling algebraic expressions (through Maxima and Sympy), as well as numerical calculations (through Scipy and Numpy), and many other tasks.

\emph{Vpython} ("visual Python")  is a 3D modeling library for Python, that uses OpenGL for rendering graphics. Its main features are its simplicity, and real-time visualization capabilities. \cite{schrerer}

\section{Equations of motion for the Bonnor metric}
Expressions for the Bonnor geodesic equations were also calculated by means of the variational principle, to contrast them with the results generated with Sage. These expressions are:
\begin{equation}
\begin{split}
\ddot{r} &=-\frac{1}{2}\frac{\partial}{\partial r}\ln\left(\frac{Y^2P^2}{Q^3Z}\right) \dot{r}^2 - \frac{\partial}{\partial \theta}\ln\left(\frac{Y^2P^2}{Q^3Z}\right)\dot{\theta}\dot{r} + \frac{Z}{2}\frac{\partial}{\partial r}\ln\left(\frac{Y^2P^2}{Q^3Z}\right)\dot{\theta}^2\\
&- \frac{Q^3Z}{2}\left[\frac{\varepsilon^2}{P^4}\frac{\partial}{\partial r}\ln\left(\frac{P^2}{Y^2}\right) + \frac{l^2}{Y^4Z\sin^2\theta}\frac{\partial}{\partial r}\ln \left(\frac{Y^2Z\sin^2\theta}{P^2}\right)\right] 
\end{split}
\end{equation}

\begin{equation}
\begin{split}
\ddot{\theta} & = -\frac{1}{2}\frac{\partial}{\partial\theta}\ln\left(\frac{Y^2P^2}{Q^3}\right) \dot{\theta}^2 - \frac{\partial}{\partial r}\ln\left(\frac{Y^2P^2}{Q^3}\right)\dot{\theta}\dot{r} + \frac{1}{2Z}\frac{\partial}{\partial \theta}\ln\left(\frac{Y^2P^2}{Q^3}\right)\dot{r}^2\\
&-\frac{Q^3}{2Y^4}\left[\varepsilon^2\frac{\partial}{\partial\theta}\ln\left(\frac{P}{Y^2}\right) - \frac{l^2}{Z\sin^2\theta}\frac{\partial}{\partial\theta}\ln\left(\frac{Y^2\sin^2}{P^2}\right)\right] 
\end{split}
\end{equation}

where the dots mean derivative with respect to $\lambda$, and $\varepsilon$ and $l$ are integration constants.

\section{Our program}
\begin{figure}
\includegraphics[width=10cm]{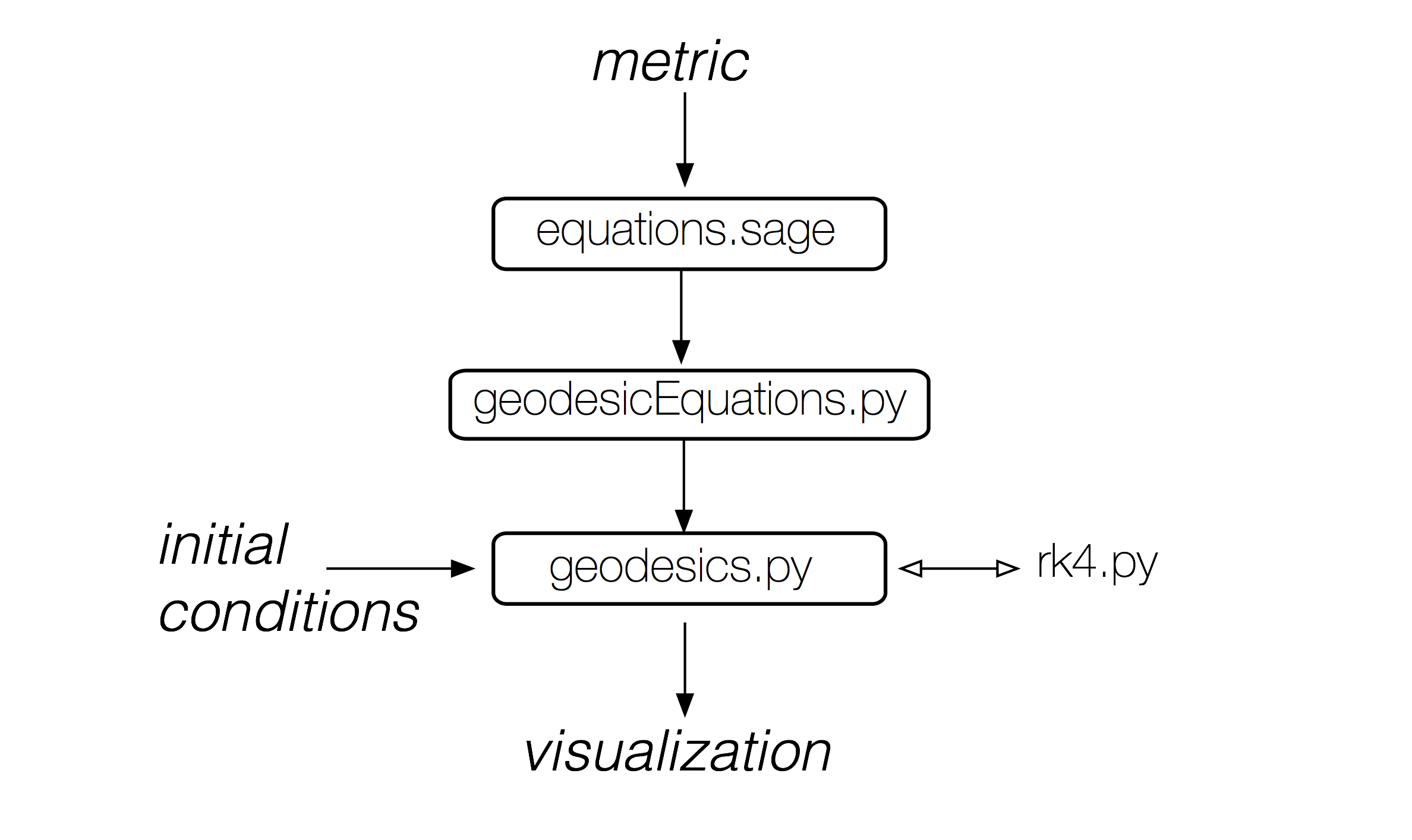}
\caption{Structure of the program}
\label{fig: program}
\end{figure}
\subsection{Analytical parts of the program}
As sketched in the figure \ref{fig: program}, the program is formed by several modules. In principle, the program is capable of taking any metric written in any coordinate system and plotting geodesics with very little changes. The first module, \verb|equations.sage|, computes the Christoffel symbols for the given metric analytically using Sage, and stores the result as a Python script that can be later used to do calculations using only Python. This Python script, named by default \verb|geodesicEquations.py|, contains the right-hand side of the Eq. \ref{eq:geodesicequation} defined as a function, as well as the metric.

\subsection{Numerical calculations}
The script \verb|geodesicEquations.py| is used in the second module, \verb|geodesics.py| to calculate and visualize the geodesics. The 4th-order Runge-Kutta method is used to solve the set of differential equations, and it is programmed in \verb|rk4.py|. The numeric calculations may also be handled using Numpy, a Python library with a variety of optimized functions written for this purpose; or a specific numerical method can also be integrated to the code this way.

The initial conditions are given to this second module, and a function is called to vary them in order to form the light beam cone. Also, these initial conditions are varied in order to get a parallel set of beams. The given initial conditions are: the initial four-vector for the position, and the three spatial initial directions \verb|v[1], v[2], v[3]| for the light beam; the other one is calculated by a function named \verb|calculate_v0()| that uses the condition that $ds^2=0$ for null geodesics. Since we are not using local frames, these spatial directions are really specified respect to a distant observer, assuming an asymptotically flat metric. Using a local frame and the condition of null geodesics, the spatial initial directions can be specified using only two angles, as done in \cite{muller}.

Users can give input to the program by creating scripts that call the files \verb|equations.sage| and \verb|geodesics.py|. In this way, multiple metrics and initial conditions can be tested with no changes to the central code, and more functionality can be added in the future.

The number of iterations and the step size can be specified, if not, default values are used. To avoid the effects of singularities on the behavior of geodesics, a stopping point based on distance to the origin is implemented and also can be specified in the main script.

\subsection{Visualization}
Vpython is used not only to plot the data, but also to manipulate the view in 3D in real time. Objects type \verb|curve()| are used for the geodesics, and two small spheres mark the positions of both the source and the origin, where the massive object is located. A distinct arrow also points in the direction of $+z$ to indicate the direction of the dipole in the case of the Bonnor metric. While it is true that the code works for any coordinate system, a transformation to the Cartesian coordinate system is needed in order to represent positions in $\mathbb{R}^3$. So far, transformations from spherical and cylindrical coordinates to Cartesian and vice-versa are already implemented.

\section{Results}

\subsection{Test runs}
\begin{figure}
\hspace*{-1.6cm}
\includegraphics[width=15cm]{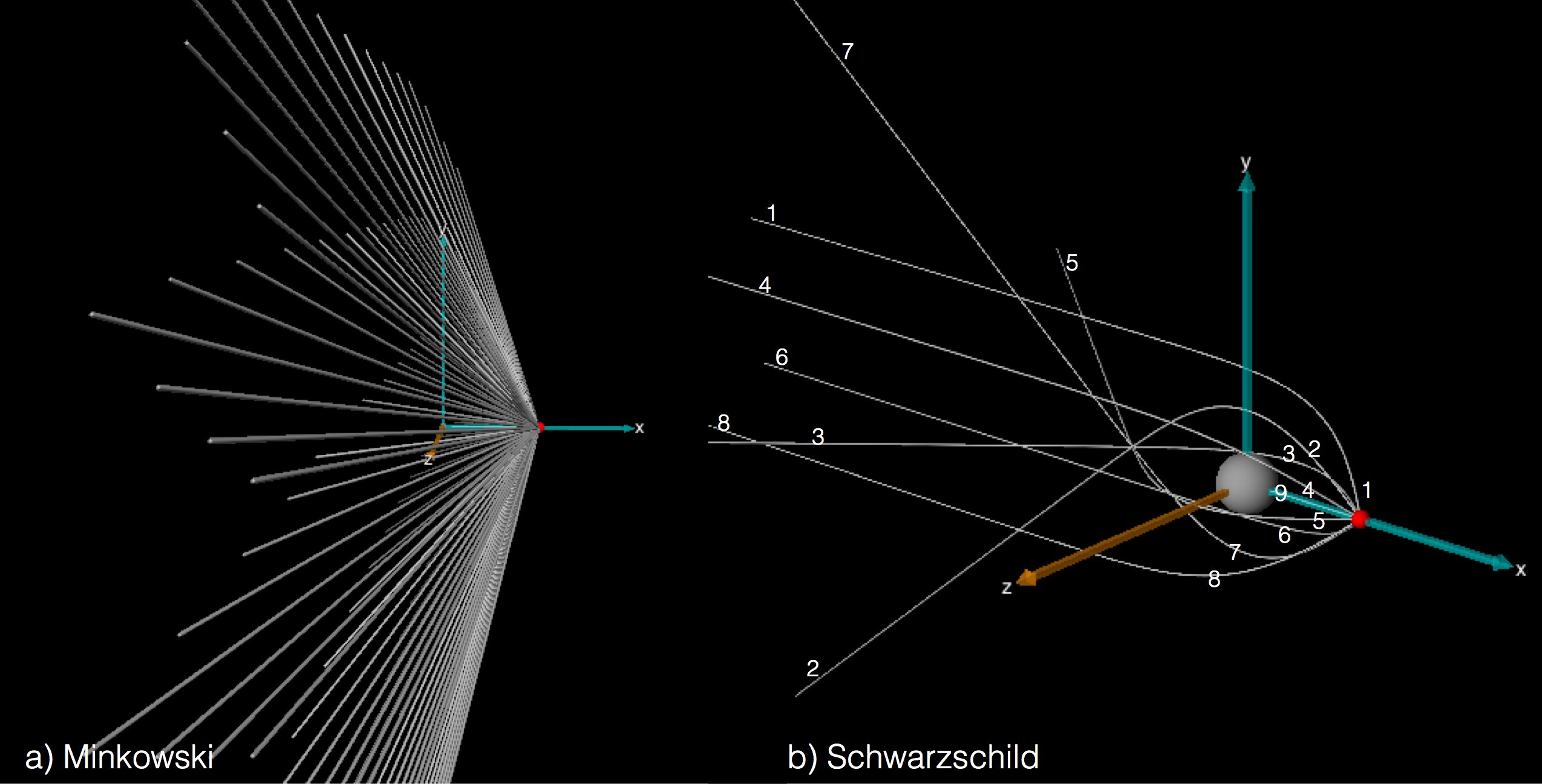}
\caption{Test runs with point sources}
\label{fig: test runs}
\end{figure}
\begin{figure}
\hspace*{-1.6cm}
\includegraphics[width=10cm]{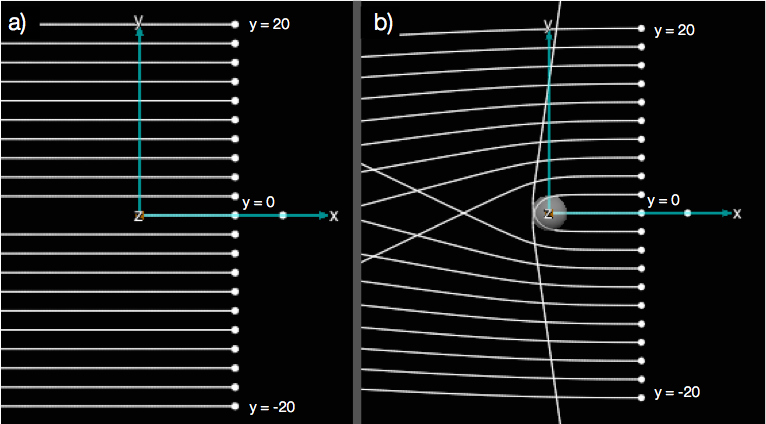}
\caption{Test runs with parallel beams}
\label{fig: b3}
\end{figure}

\begin{figure}
\hspace*{-1.6cm}
\includegraphics[width=17cm]{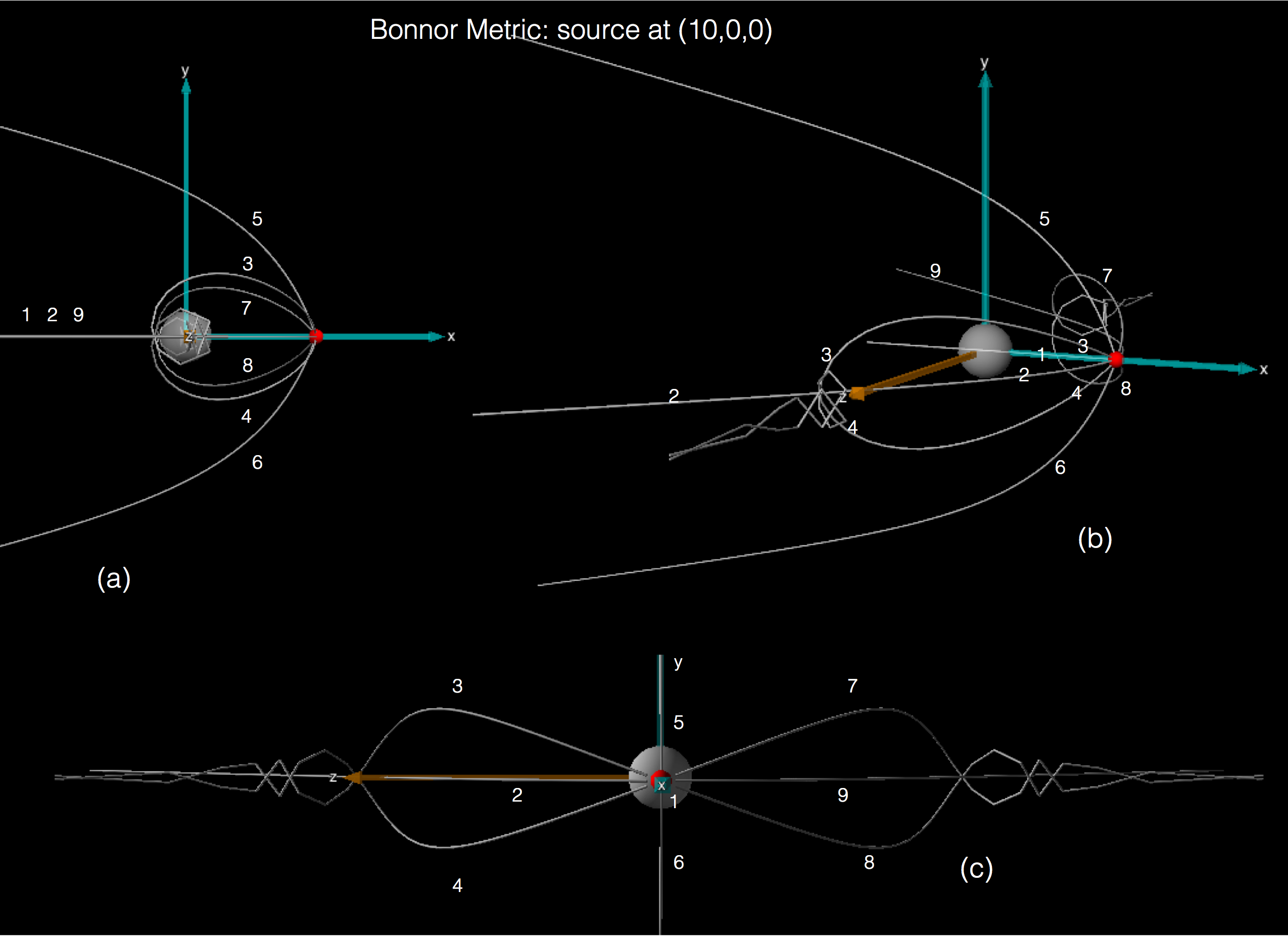}
\caption{Geodesics for the Bonnor metric with $b=0.001$}
\label{fig: b1}
\end{figure}

The program was tested first with Minkowski and Schwarzschild metrics. In the Fig. \ref{fig: test runs}a a light beam cone is drawn in a Minkowski space-time (in spherical coordinates), presented here to compare with the results obtained with other metrics. As it can be seen, each geodesic starts at the same point, marked by a small red sphere in the figure. As expected, geodesics are straight lines. Results for Schwarzchild metric are shown in Fig. \ref{fig: test runs}b. Fewer geodesics are displayed this time to better show how they are curved by the space-time. As expected, the light beams even cross each other when passing very close to the event horizon, and they are just slightly deviated if they pass farther away. 

Now, parallel beams are simulated for both metrics, and the results are shown in Fig. \ref{fig: b3}. In both cases, the direction of the beams is the negative $x$ axis, and all the geodesics remain in the $x$-$y$ plane. The mass for the Schwarzschild metric is set to 0.5.

The main scope for running these test metrics, where geodesics are known, was to detect possible errors and select more carefully the initial conditions for later runs.

\subsection{Point source with $b<1$}
Now, to the Bonnor metric. In Fig. \ref{fig: b1}, nine geodesics have been simulated and viewed from different angles, to give a sense of spatial orientation. The source is located\footnote{These position vectors are given in Cartesian coordinates; units are in the geometric system.} at $(10,0,0)$, and the direction of the central geodesic is radially inwards, varied then in $-0.8\leq \verb|v[2]| \leq 0.8$ and $-0.3\leq \verb|v[3]| \leq 0.3$, corresponding to the $\theta$ and $\phi$ directions. 20 iterations with \verb|dlam = 0.1| (step size) have been carried out. The metric parameters are $m=1$ and $b=0.001$.

 The presence of the magnetic field makes the photons move in an helical motion around the $z$ axis, that is, around the poles of the magnetic dipole, forming a jet-like structure. This behavior is consistent to the behavior of null geodesics for the Melvin metric, that is, locally, the magnetic field is dominant (see \cite{bonatti1}).
 
 \begin{figure}
 \hspace*{-1.6cm}
 \includegraphics[width=14cm]{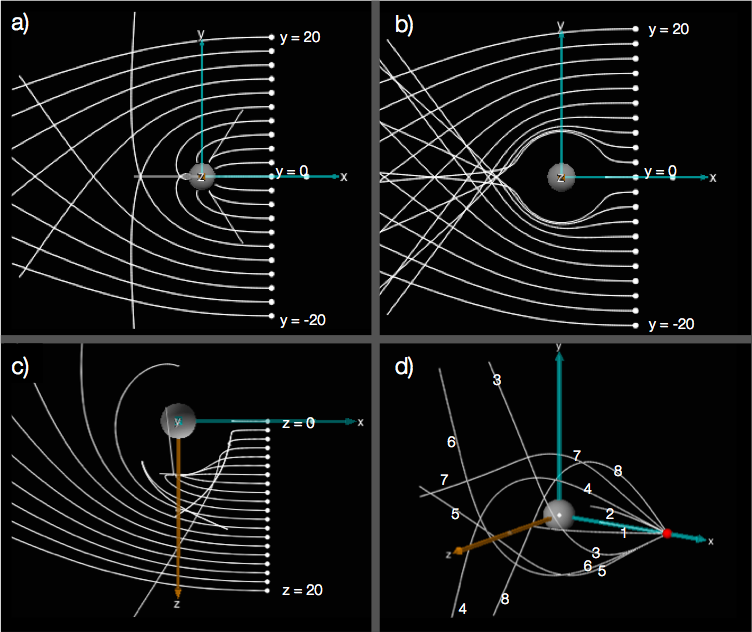}
 \caption{Geodesics for the Bonnor metric}
 \label{fig: b4}
 \end{figure}
 
 \subsection{The values of $b$ and $m$}
Various values of $b$ and $m$ were tested. The behavior obtained in the preceding section corresponds to tested values $1\leq b \leq 10^{-30}$. With values of $b$ in the interval $2\leq b\leq 10$ at least, the swirling seems to disappear. As predicted by Bonnor, with a value of $m=0$ the space-time becomes flat and the result is similar to Fig. \ref{fig: b3}a. With a value of $b=0$, also pointed out by Bonnor, the metric does not tend to the Schwarszchild one, but can be interpreted as a monopole of mass $2m$ together with higher mass multipoles dependent on $m$. Parallel beams of light on both metrics show the difference (compare Fig. \ref{fig: b4}a and Fig \ref{fig: b3}b).

\subsection{Light beams with $b\sim 5$}
For the tested values of $b=2,5,10$ and $m=1$ the results were similar. In Fig. \ref{fig: b4}b, the same $x$-$y$ view is shown, for comparison with Figs. \ref{fig: b4}a and \ref{fig: b3}b. The magnetic field bends the light in such a way that a big opening is created around the object, and the light seems not to collide with it. Now, in Fig. \ref{fig: b4}c, variation in the $z$ axis is tested to see the effect on the distance for the magnetic dipole. As expected, the effect decreases with distance, even in the $z$ axis where the dipole points to. It is important to notice that for Figs. \ref{fig: b4}b and \ref{fig: b4}c, the curves shown are all in the same plane.

A point source located at $(15,0,0)$ shows a different story (Fig. \ref{fig: b4}d). Geodesics are no longer necessarily contained in a plane, and bend creating an "envelope" around the object that is less evident with the distance.

\section{Physical remarks}
Parallel beams of light simulate distant sources and therefore can be used for gravitational lensing, helping to diagnose which kind of object causes a particular deformation on an image. Neutron stars can have very strong magnetic fields, up to the point of not being ignorable by General Relativity, but also have very rapid rotation rates, usually dominant. Light that behaves as shown in this paper, then, alerts of the presence of a massive object with a very strong magnetic field, but with negligible rotation. A point source, in the other hand, is also important to take into account, since several of these massive objects (black holes or neutron stars) also have companions, that is, nearby stars that act like point sources of radiation.

\end{document}